\documentclass[onecolumn]{article}
\usepackage{graphicx}

\newcommand{\newsection}[1]{
\addtocounter{section}{1}
\setcounter{equation}{0}
\setcounter{subsection}{0}
\addcontentsline{toc}{section}{\protect
\numberline{\arabic{section}}{{\rm #1}}}
\vglue .6cm
\pagebreak[3]
\noindent{\bf  \thesection. #1}\nopagebreak[4]\par\vskip .3cm}
\newcommand{\newsubsection}[1]{
\addtocounter{subsection}{1}
\addcontentsline{toc}{subsection}{\protect
\numberline{\arabic{section}.\arabic{subsection}}{#1}}
\vglue .4cm
\pagebreak[3]
\noindent{\it \thesubsection. #1}\nopagebreak[4]\par\vskip .3cm}

\newcommand{\ben}{\begin{enumerate}}
\newcommand{\een}{\end{enumerate}}

\newcommand{\be}{\begin{equation}}
\newcommand{\ee}{\end{equation}}
\newcommand{\bea}{\begin{eqnarray}}
\newcommand{\eea}{\end{eqnarray}}

\begin{document}
\begin{flushright}
\end{flushright}
\vspace{0.1cm}
\thispagestyle{empty}

\begin{center}
{\Large\bf 
The close-packed triple helix as a possible new structural motif for collagen}\\[13mm]
{\rm  Jakob Bohr\footnote{jakob.bohr@fysik.dtu.dk} and Kasper Olsen{\footnote{kasper.olsen@fysik.dtu.dk}}}\\[2.5mm]
{\it Department of Physics,\\ Technical University of Denmark}\\
{\it Building 307 Fysikvej, DK-2800 Lyngby, Denmark}\\[6mm]
\end{center}

\begin{abstract}
The one-dimensional problem of selecting the triple helix with the highest volume fraction is solved and hence the condition for a helix to be close-packed is obtained. The close-packed triple helix is shown to have a pitch angle of $v_{CP} =43.3 ^\circ$. Contrary to the conventional notion, we suggest that close packing form the underlying principle behind the structure of collagen, and the implications of this suggestion are considered. Further, it is shown that the unique zero-twist structure with no strain-twist coupling is practically identical to the close-packed triple helix. Some of the difficulties for the current understanding of the structure of collagen are reviewed: The ambiguity in assigning crystal structures for collagen-like peptides, and the failure to satisfactorily calculate circular dichroism spectra. Further, the proposed new geometrical structure for collagen is better packed than both the $10/3$ and the $7/2$ structure. A feature of the suggested collagen structure is the existence of a central channel with negatively charged walls. We find support for this structural feature in some of the early x-ray diffraction data of collagen. The central channel of the structure suggests the possibility of a one-dimensional proton lattice. This geometry can explain the observed magic angle effect seen in NMR studies of collagen. The central channel also offers the possibility of ion transport and may cast new light on various biological and physical phenomena, including biomineralization.
\end{abstract}
\newpage
\tableofcontents

\newsection{Introduction}
\label{intro}

A broad range of materials with various types of molecular forces form close-packed atomic structures \cite{burns1985,finnis2004,buckingham1993}. Of the elements, the noble gases  and many metals from close-packed crystal structures, i.e. a significant fraction of the periodic system. For simple atomic lattices, close-packing is defined by the geometry of hard spheres which minimize the volume occupied by the spheres. Likewise, helical structures formed of flexible tubes with hard walls have a most favorable packing in the sense of the best use of space.  Recently, we have shown that the helical motifs of DNA and alpha-helices are in rather fair agreement with the close-packed double and single helices \cite{olsen2009}. This opens the question of the agreement not being coincidental, and suggests that the optimization of the volume fraction may be at play as a contributing factor in the formation of the molecular structures. 

Some structural motifs of biological chain molecules have previously been discussed as various conformations of tubes \cite{banavar2007}. When emphasizing one kind of optimum shape of helices, knots and other structures, the global curvature of a space curve has been introduced, see Gonzalez and Maddocks \cite{gonzalez1999} and Maritan et al. \cite{maritan2000}. As an example of such an optimum structure, Maritan et al. \cite{maritan2000} depict the helical structure with a pitch angle of $v_{TP}= 21.8^\circ$. This helical structure is at the transition between two regimes \cite{maritan2000}; in ref. \cite{olsen2009}  we have denoted this particular structure as being {\it tightly packed}. In the case of the single helix, the pitch angle of the tightly packed helix differs only by a relatively small amount from that of the close-packed helix. However, for double helices the difference in the pitch angle for the two structures is $12.5^\circ$. Moreover, one of the interesting features of the close-packed double helical structure is that it has a central channel (the tightly packed double helix has no central channel). 

In this paper, we will investigate the packing of the triple helix and thereby revisit the question of the structure of collagen. The phrase {\it triple helix} is ambiguous; some of us will instinctively think of three interwoven helices that share the same axis around which they are twisted;  others will think of three (single) helices which are twisted individually around their own axis and which may be 
super-coiled (SC) around each other. The first category is akin to the double-helix of DNA, while the second type is akin to the super-coiling of two 
$\alpha$-helices. 

Ramachandran \& Kartha \cite{ramachandran1954,ramachandran1955} had the insight to suggest a triple helix for the structure of collagen in 1954. This insight became supported by Rich and Crick \cite{rich1955,rich1961} who suggested a variant - known as the super-coiled 10/3 structure. The triple-strand nature of the structure was corroborated by sequencing. Later, Okuyama et al. suggested the super-coiled 7/2 structure \cite{okuyama1977}.  Presumably, the long collagen triple helix can be incommensurate, a view accredited to have been shared by Ramachandran \cite{sasisekharan1999}, see also Cameron et al. \cite{cameron2007}. The supermolecular structure of fibrous collagen has been examined in a synchrotron x-ray study by Orgel et al. \cite{orgel2006}. Collagen-like peptides have been theoretically investigated by DFT \cite{tsai2005} and molecular dynamics \cite{punitha2009}. For a review of the structure and stability of the long collagen molecules and the fibrils, see Kadler et al. \cite{kadler2007} and Shoulders and Raines \cite{shoulders2009}. For a discussion of mechanical properties on different length scales see Buehler \cite{buehler2006}.

Interestingly, it is still debated whether the $10/3$ structure (3.3 residues/turn) suggested by Rich and Crick \cite{rich1955} or the $7/2$ structure (3.5 residues/turn) suggested by Okuyama et al. \cite{okuyama1972,okuyama1977} is the better descriptor of collagen? Some collagen-like peptides have been assigned the $10/3$ structure in crystallographic structures, see Bella et al. \cite{bella1994}, for a summary see Beck and Brodsky \cite{beck1998}. Other peptides have been assigned the 7/2 structure \cite{okuyama1999}. Descriptions with variable twist angles are now common \cite{berisio2002,okuyama2004,okuyama2006a,okuyama2006b}. In all cases, the R-factors have remained relatively large. To us it leaves open the possibility that a third structure would be in better agreement with the data than either of the ones that were considered in the past.

\newsection{The triple helix}
\label{sec:2}

\begin{figure}[h]
\begin{center}
\includegraphics[width=7.5cm]{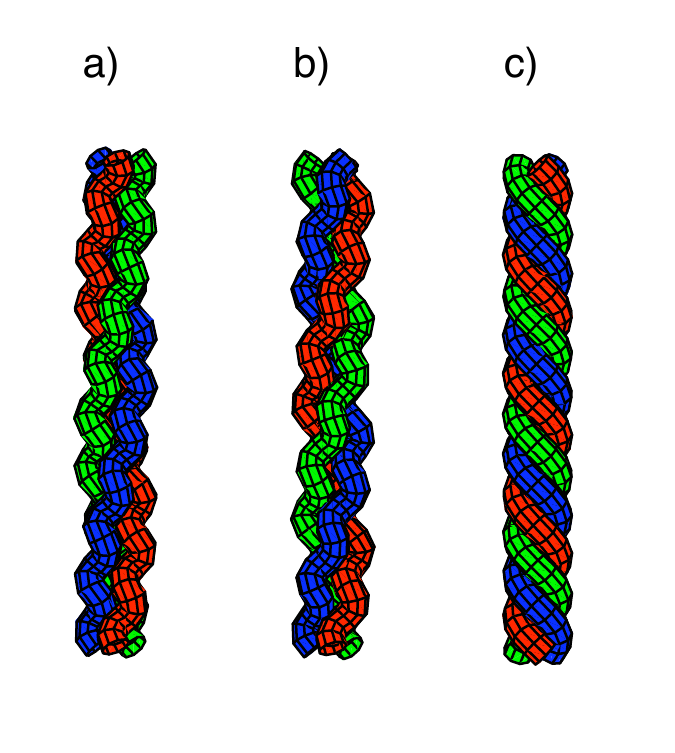}
\caption{{\it Different topological structures of triple helices made out of three identical tubes of tube diameter $D$.  a) Supercoiled 10/3 structure. b) Supercoiled 7/2 structure. c) The close-packed (CP) triple helix (left-handed variant), which is not supercoiled}}
\end{center}
\end{figure}

We suggest the close-packed tube geometry as a possible packing principle of the triple helical structure of collagen based on the knowledge of its higher volume fraction. Here we develop the motif in some detail. Figure 1 a) and b) shows the triple helices as three single helices that are super-coiled around each other. Figure 1 c) shows the close-packed motif that consists of three interwoven helices that share the same axis. For the triple-helix with a single twist axis the parametric equations are 

\begin{eqnarray}
\vec{r}_1 &=& (a\cos t_1, a\sin t_1, ht_1) \\
\vec{r}_2&=&(a\cos t_2, a\sin t_2, ht_2+\frac{2\pi h}{3})\\
\vec{r}_3&=&(a\cos t_3, a\sin t_3, ht_3+\frac{4\pi h}{3})
\end{eqnarray}
where $h=H/2\pi$ is the reduced pitch height and
$a$ is the radius of the cylinder surface hosting the helical curves. The direction of the tangent is found through differentiation, 
$d\vec{r}_i/dt_i=(-a\sin t_i, a\cos t_i, h)$.
The tangent to each of these curves is at the helix angle, $v$, with the vertical axis. Hence, the pitch angle, $v_{\bot} =90^{\circ} - v$ is determined by
\begin{equation}
\label{ }
\tan v_\bot = \frac{h}{a}~.
\end{equation}

\noindent The triple helical lines presented above and the analyses in the following follow a right-handed helical structure. The optimum for a left-handed system is simply found by mirror symmetry and therefore imposes the same scale parameters on the triple-helix. In Figure 1c) is depicted the left-handed solution which is also the one chosen for the atomistic considerations below as it appears to have a more realistic Ramachandran plot.

For two points on the surface of the helical tubes to be in contact, it is a requirement that the distance between the central helical lines of the tubes is a minimum. Consider two points on two of the helical lines of the symmetric triple-helix:
$\vec{r}_1 = (a, 0, 0) $ and 
$\vec{r}_2=(a\cos t, a\sin t, h t+2\pi h/3)$.
The square of their distance, $D_3$, is
\begin{equation}
\label{ }
D_3^{2}=|\overrightarrow{r_1r_2}|^2 = a^2(\cos t-1)^2 + a^2\sin^2 t  + (\frac{2\pi h}{3}+ht)^2;
\end{equation}
the derivative determines the possible minima. 
We find the condition that $D_3$ is an extremum as:
\begin{equation}
\label{transient}
\sin  t+\frac{h^2}{a^2}\frac{2\pi}{3 } +\frac{h^2}{a^2}t=0~.
\end{equation}
The triple-helix is packed when $D_{3} = D$, where $D$ is the diameter of the tubes. For discussions of contacts in tube models see also Przyby\l~and Piera\'{n}ski \cite{przybyl2001}, and Neukirch and van der Heijden \cite{neukirch2002}.

\newsubsection{The close-packed triple helix}

The close-packed triple helix is found by considering an enclosing cylinder of volume $V_E=2\pi^2 h(a+D/2)^2$, and comparing it to the volume occupied by the three helical tubes, $V_H = 3\pi H D^2 /(4  \sin v_\bot)$. The volume fraction is the ratio $V_H/V_E$:

\begin{equation}
\label{}
f_V 
= 3 (1+(\frac{a}{h})^2)^{1/2} \cdot (\frac{2a}{D}+1)^{-2}~.
\end{equation}

\noindent 
In Figure 2 is shown $f_V$ for the packed helices. Numerically, the maximum is found to be at a pitch angle of $v_{CP} = 43.3^\circ$, and the corresponding fraction of volume occupied to be $f_{CP}=0.744$. The unique packed structure with this value of $v_\bot$ is henceforth called the close-packed (CP) triple helix.
\begin{figure}[t]
\begin{center}
\includegraphics[width=7.5cm]{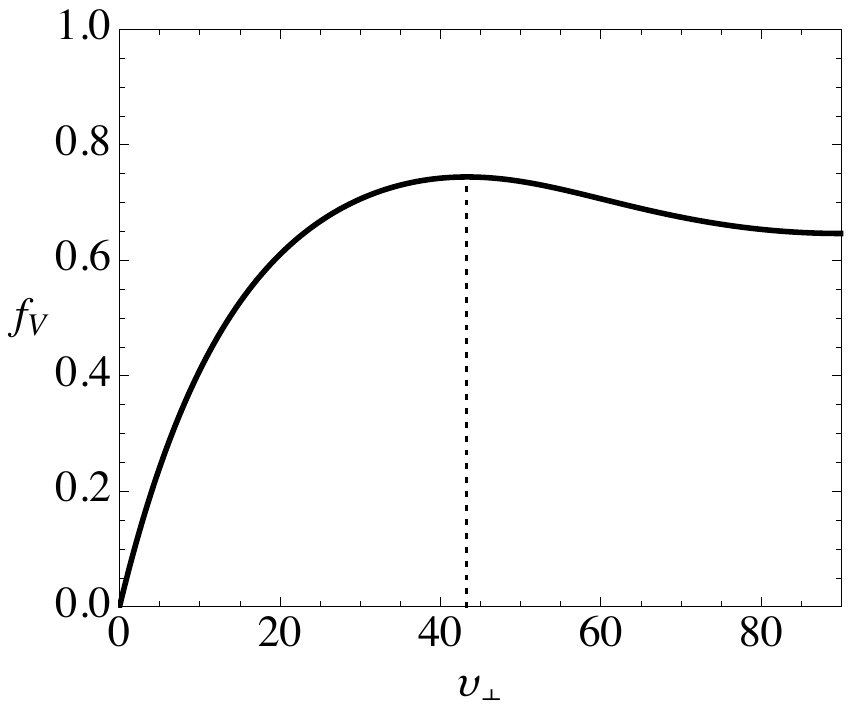}
\caption{{\it Volume fraction $f_V$ for the packed triple helix. The maximum packing fraction is $f_{CP}=0.744$ and is obtained for a pitch angle $v_{CP}=43.3^\circ$. The dashed vertical line represents the corresponding close-packed (CP) structure}}
\end{center}
\end{figure}

For the volume of the tube models in Figure 1 to be compared, the scale of the parameters needs to be determined. As the diameter of the three backbone tubes we use $D_{p}=5.0$ \AA~ estimated from the tube model of the $\alpha$-helices ($a=2.3$ \AA ~and $2a/D_{p}=0.91$) \cite{olsen2009}. For the raise per residue, $H_R$, and per triplet of residues, $H_T = 3 H_R$, we will use 2.86 \AA ~and 8.57 \AA, numbers well-known from x-ray diffraction \cite{beck1998,okuyama1999}. These numbers are the same for the $10/3$ and the $7/2$ models also  discussed. For the super coiled structure, the corresponding volume fraction determined by an enclosing cylinder is found to be about 0.6. This number depends on how the single helices are rotated relative to the super coiled structure. It is $\sim$ 20\% smaller than the volume fraction (0.74) of the close-packed structure.\\

\newsubsection{The central channel}

\begin{figure}[h]
\begin{center}
\includegraphics[width=7.5cm]{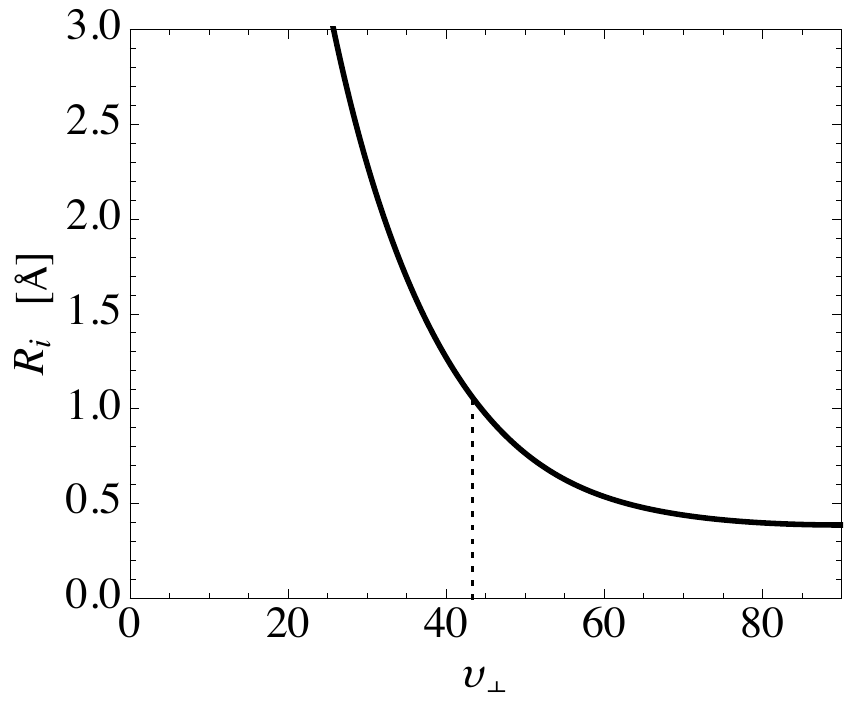}
\caption{{\it All the packed triple helices have a central channel with non-vanishing inner radius $R_i$. 
The minimum values of $R_i$ are here plotted as a function of pitch
angle, $v_{\bot}$, for the triple helix with tube diameter $D=5$ \AA. The close-packed triple helix, dashed vertical line at $v_\bot=43.3^\circ$, has an inner radius of $R_i= 1.06$ \AA}}
\end{center}
\end{figure}
It is a feature of the packed triple helices that they all have central channels. This is not the case for packed double helices where no central channel exists for $v_\bot \geq 45^\circ$ \cite{olsen2009}. The radius, $R_i=a-D/2$, of the central channel increases with decreasing pitch angle, $v_\bot$, see Figure 3. The numbers which characterize the close-packed structure are given in Table 1. 
\begin{table}[h]
\caption{{\it The close-packed (CP) triple helix:  $a$ is the radius of the hosting cylinder, $v_{\bot}$ is the pitch angle, $D$ is the diameter of the three tubes, $R_i$ the radius of the central channel, and $R_o$ the radius of a cylinder circumscribing the triple helix; $f_V$ is the packing fraction. The bottom row applies to the proposed collagen motif.} }
\label{tab:1}
\begin{tabular}{lllllll}
\hline
Type &  $a$ [\AA] &$v_\bot$ [$^\circ$] &  $D$ [\AA]&$R_i$ [\AA]&$R_o$ [\AA]&$f_V$\\
\noalign{\smallskip}\hline\noalign{\smallskip}
CP & $a$ & 43.3 &  1.404 $a$& $a-D/2$ & $a+D/2$ & 0.744\\
CP & 3.56 & 43.3  & 5.0 & 1.06 & 6.06 & 0.744\\
\hline
\end{tabular}
\end{table}

\newsubsection{The zero-twist structure}

\begin{figure}[h]
\begin{center}
\includegraphics[width=7.5cm]{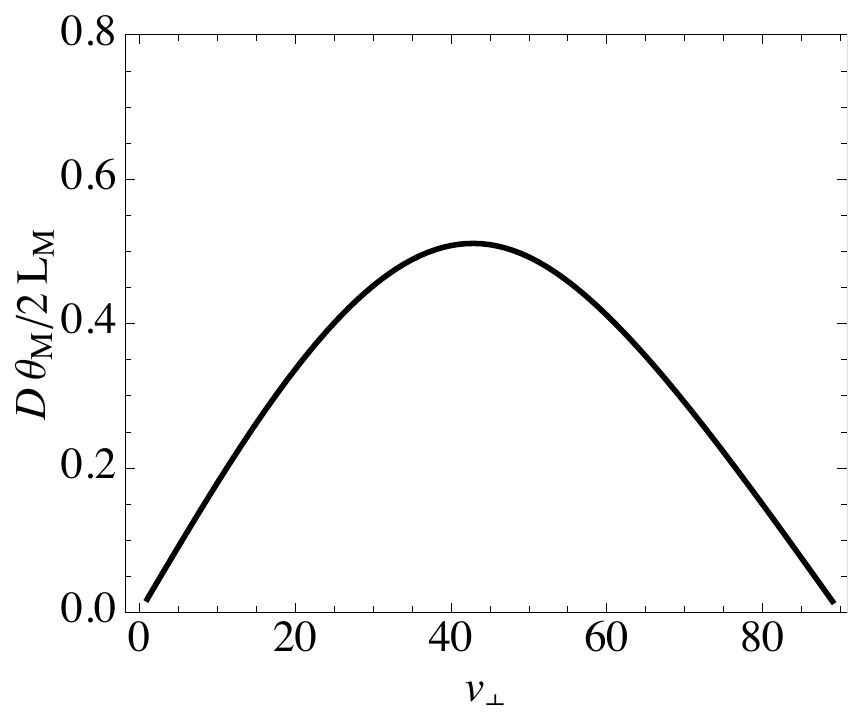}
\caption{{\it The dimensionless normalization, $D\theta_M/2L_M$, of the twist, $\theta_M$, of one of the three tubes (length $L_M$) plotted as a function of the pitch angle. It is assumed that the helices are packed, i.e. that they maintain self-contact. The maximum value of the twist appears for $v_{ZT}=42.8^\circ$. At this pitch angle there is no geometrical coupling between strain and twist}}
\end{center}
\end{figure}

Generally, helical structures have a significant strain-twist coupling. It follows straightforwardly, that a straining (stretching) of a helix leads to a change in its total twist, as the length of a helical structure is affected by how the helix is twisted. If a helix is being strained, some of the stress can be released by a small change in the pitch angle. For a long molecule the consequence is that one end may rotate by a significant amount relative to the other end. Here, we investigate this in more details. The total twist, $\theta_M$, of one of the helical tubes, as one advances the length $L_M$ along the tube is

\begin{equation}
\label{}
\theta_M = \frac{L_M \cos v_{\bot}}{a}~.
\end{equation}

\noindent For the triple helix, $\theta_M/L_M$ has its maximum at $v_\bot=42.8^\circ$, see Figure 4. At this particular angle there is no differential change in $\theta_M$ with $v_{\bot}$ and therefore no geometrical coupling of strain to twist. We denote the corresponding structure as the {\it zero-twist} structure. The zero-twist angle, $v_{ZT}=42.8^\circ$, is essentially equal to the pitch angle for close-packing, $v_{CP}=43.3^{\circ}$. This is a unique feature of the triple helix which does not hold for the double helix, nor for the quadruple helix, see Table 2. The CP triple helix motif is therefore particularly well-suited to form very long structures in situations where a zero-twist structure is desirable. E.g. in tendon and bone.\\ 

\begin{table}[h]
\caption{{\it The pitch angles of the zero-twist structures, $v_{ZT}$, and of the close-packed helices, $v_{CP}$, for the double, triple, and quadruple helices.} }
\label{tab:2}
\begin{tabular}{llll}
\hline
No. of strands &  $2$&$3$ & $4$\\                        
\noalign{\smallskip}\hline\noalign{\smallskip}
$v_{ZT}$ & $39.4^\circ$ & $42.8^\circ$ &  $43.8^\circ$\\
$v_{CP}$ & $32.5^\circ$ &$ 43.3^\circ$  &$53.7^\circ$\\
\hline
\end{tabular}
\end{table}
\vspace{2cm}

\newsection{Discussion}

Here we review some of the implications if collagen were packed in accordance with the CP triple helix motif.

\newsubsection{Equatorial scattering and the Haisa data}

In order to investigate if scattering of x-rays in the equatorial plane could distinguish 
between the CP and the SC structures the squared structure factors are calculated as

\begin{equation}
|S(\vec{q})|^2\propto |\int  \rho (\vec{r}) \exp (-i \vec{q} \cdot \vec{r}) dA |^2
\end{equation}
where $\vec{q}$ is the scattering vector and $\rho$ the electron density.
The projection of $\rho$ on the equatorial plane depends only on $r$. If we assume there is water with $\rho=1$ surrounding the structure, we can calculate $|S(q)|^2$ more easily by subtracting a constant of 1 from $\rho$. The absolute square of the structure factor becomes

\begin{equation}
| S(q) |^2 \propto |\int_0^{R_o} \int_0^{2\pi} \rho_e (r) \exp(-i q r \cos \phi)  r d\phi d r|^2,
\end{equation}

\noindent where $\rho_e (r) = \rho(r) -1$. Figure 5 shows the calculated density profiles, $\rho_e(r)$, and intensity profiles $|S(q)|^2$ for four different cases a) the CP structure with channel width 1.6 \AA ~, b) box density profile with central channel, c) Tube structure without channel akin to the two SC structures, and d) box density profile. Only the structures with a channel show appreciable intensity at wave vectors from about 1 to 2.5 \AA$^{-1}$.

\begin{figure}[h]
\begin{center}
\includegraphics[width=8.5cm]{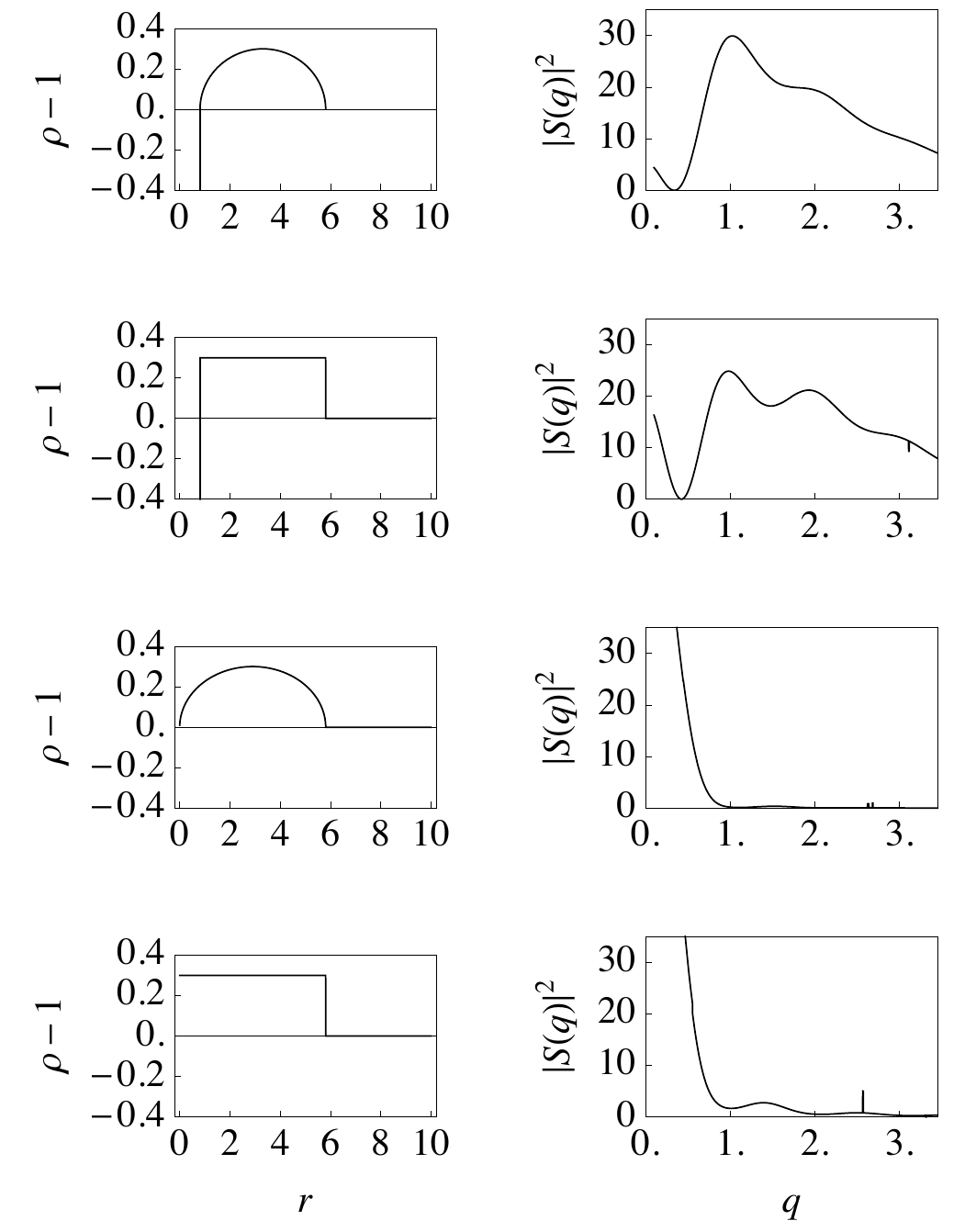}
\caption{{\it Calculated density profiles, $\rho_e(r)$ (left column), and intensity profiles, $|S(q)|^2$ (right column) for four different cases: a) the tube structure with a channel of diameter 1.6 \AA ~, b) box density profile with central channel of the same diameter, c) tube structure without channel akin to the two super-coiled structures, and d) box density profile with no central channel. Here $r$ is measured in \AA, $q$ in \AA$^{-1}$, and $|S(q)|^2$ in arbitrary units}}
\end{center}
\end{figure}

The CP helical structure has not previously been discussed in the literature. It is therefore necessary to carefully reassess if existing x-ray data give support for the CP structure. We find the early paper of Haisa \cite{haisa1962} to be useful as it presents absolute intensity as a function of the scattering angle in the range of interest. The paper report on changes in the diffraction pattern upon various chemical treatments of collagen. In Figure 6 we have redrawn the data for native moist collagen as a function of the scattering vector. As one can see, there is a strong resemblance of the diffraction pattern from about 1 \AA $^{-1}$ and higher to the 
Figures 5 a) and b). The peak in the diffraction data at about 0.38 \AA $^{-1}$ stems from interference from one triple helix to a nearby one. Its distance can be estimated to be approximately $4 \pi /q\sqrt{3} = 19$ \AA. We take the Haisa data as a hitherto overlooked experimental indication for the existence of a central channel in collagen. The published 2d-diffraction film images of Ramachandran \cite{ramachandran1954}, and Okuyama et al. \cite{okuyama1972} display equatorial streaks of intensity characteristic of a sharp density variation with continuous symmetry in the longitudinal direction.

\begin{figure}[h]
\begin{center}
\includegraphics[width=7.5cm]{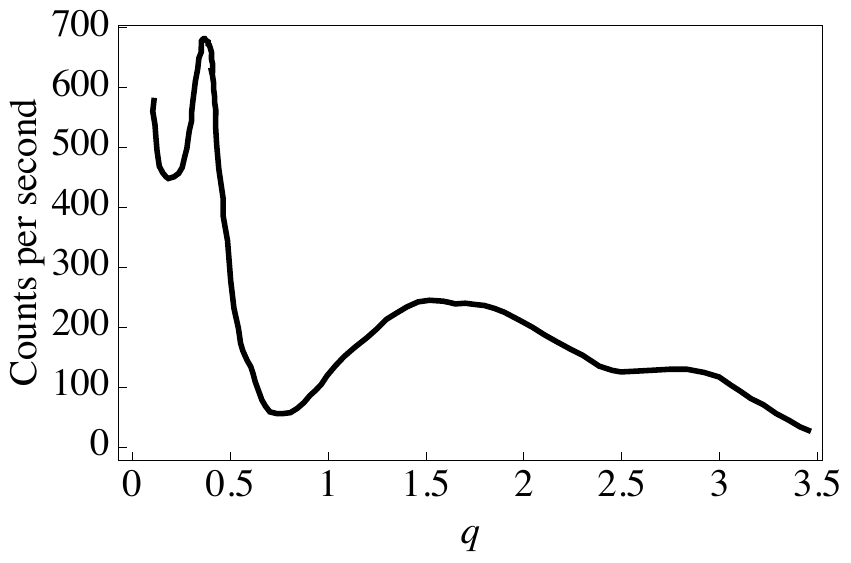}
\caption{{\it Intensity as a function of the scattering vector, $q$, in X-ray scattering from native, moist collagen. 
The original data from Haisa \cite{haisa1962} were plotted as a function of scattering angle. Here, after digitizing we have replotted the data as a function of scattering vector. As one can see there is a resemblance of the diffraction pattern from about 0.5 \AA $^{-1}$ and higher to the 
Figures 5 a) and b). The peak in the diffraction data at about 0.38 \AA $^{-1}$ stems from interference from one triple helix to a nearby one}}
\end{center}
\end{figure}

\newsubsection{Crystallographic data}

Significant interest in studying collagen with crystallographic methods has been executed \cite{okuyama2008}. For such studies, collagen-like peptides typically 30 residues long has been used.

The structure of collagen-like peptides  have been modeled by supercoiled helices (described above), in particular by the 7/2 and 10/3 structures . Here, the 7/2 structure mean that the period of the supercoiling is seven triple residues long and for the 10/3 structures the supercoiling is ten triple residues long, i.e. 30 residues. The obtained R-factors are high, typically above 20 \%.

The backbone line of a super-coiled (SC) structure can be described by the following center curve, $\vec{r}(t)$:
\begin{equation}
(a_s \cos (\frac{t}{h_s}) +a_p \cos (\frac{t}{h_s} +\frac{t}{h_d}), a_s \sin (\frac{t}{h_s}) +a_p \sin(\frac{t}{h_s} +\frac{t}{h_d}),t)
\end{equation}
where 
\begin{equation}
h_d=\frac{h_s h_p}{h_s-h_p}~.
\end{equation}
If $a_s$, or $a_p$, are zero we have the simple helix such as in CP structures, otherwise we have a supercoiled structure. For some CP-like structures one might expect a small value for $a_s$ rather than zero as this is the simplest kind of symmetry break the structure can make. 
However, the two structures will not convert from the one to the other with the use of a typical refinement algorithm. As can be noticed, if transformations are made between a SC  structure (e.g. 7/2 structure) and a CP, or near CP, structure there will be intermediate structures where the density of atoms in the center of the molecule will exceed unity by a multiple factor (the three backbones need to cross through each other). This will constitute a barrier for refinement and it is therefore highly unlikely that structure refinements which started with a SC structure could be able to relax to a CP structure.

\newsubsection{Relationships between atomistic SC and CP structures}

\begin{figure}[t]
\begin{center}
\includegraphics[width=9.5cm]{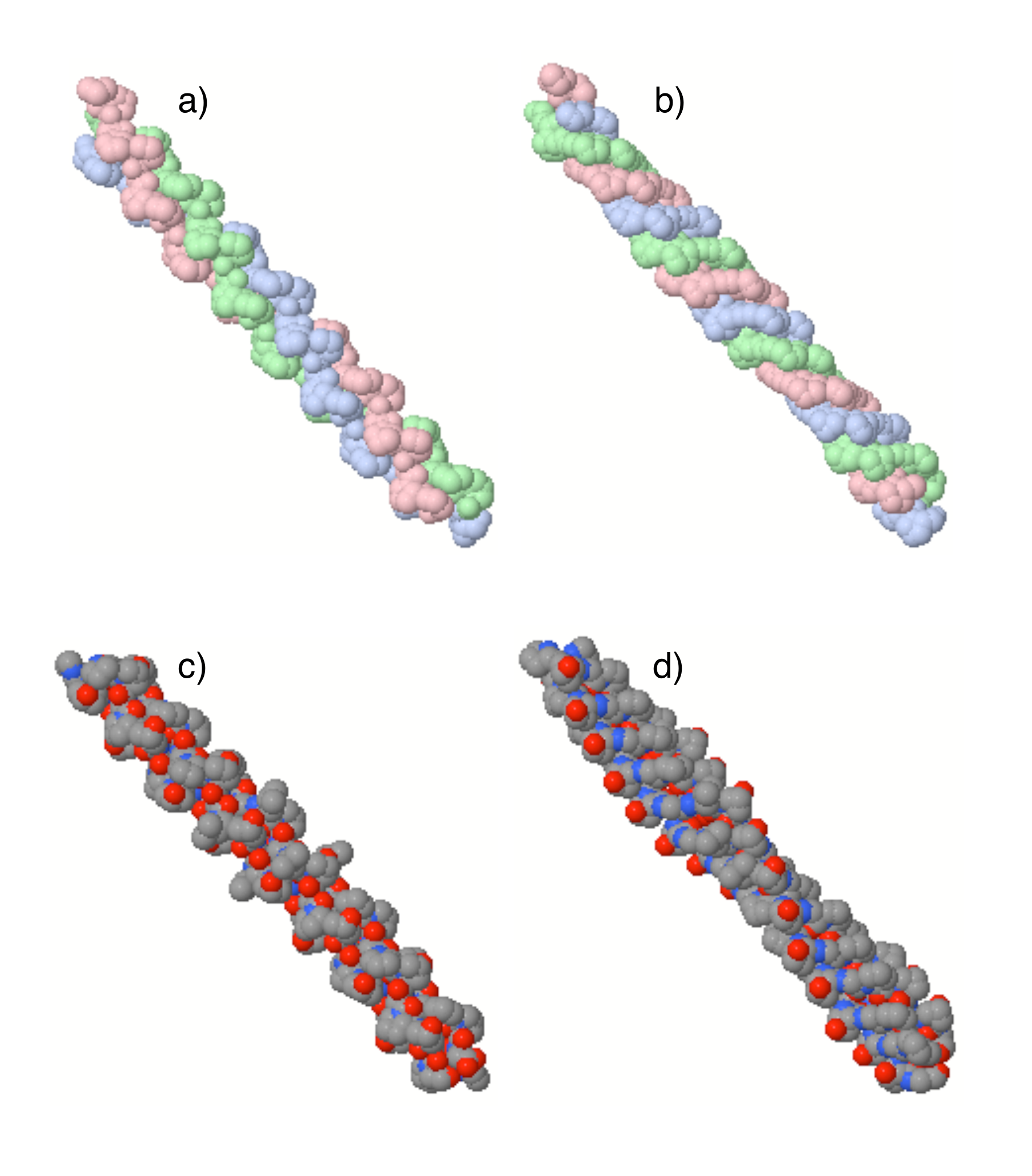}
\caption{{\it Approximate atomistic models and PDB entries: a) of the PDB 1k6f structure, and b) of the alternative CP structure derived in this paper, c) of the PDB 2drx structure, and d) of the alternative CP structure. Notice the helical lines  of the Hydroxyproline OH groups in d)}}
\end{center}
\end{figure}

Here we investigate the relationship between the CP and SC structures depicted in Figure 1 and reflect on their plausibility. 

[Pro-Pro-Gly]$_{10}$: As a first attempt to create an atomistic CP model, we build on the relationship between the CP structure and the SC 7/2 and 10/3 structures. We start by the simple and classical Pro-Pro-Gly collagen-like peptide \cite{berisio2002}. Its PDB entry 1k6f is in the SC motif, depicted in Fig. 7 a). This structure has no side-chains and is therefore particularly symmetric. For the purpose of bringing the three peptide chains in their right positions, we first deconstruct the supercoiled structure and then construct the CP structure in the following way: we unwind the supercoil, and thereafter unwind each of the three peptide helices. Thereafter, the rotational phases of each of the three chains and the radial position of the chains are adjusted, and finally the three strands are coiled to the CP structure. In Figure 7 b) we have depicted the resulting atomic structure. This method does not maintain all bond lengths. In particular, the individual nature of the refined atomic positions in the PDB entry for the SC structure transform into a kind of random noise for the reconstructed CP structure. Nevertheless, one can get an impression of the main features of the derived structure and of the relationship between the structures. The CP structures are not refined by force fields or by other means. In the presented structure there is some freedom for each of the strands to rotate around their own axis, see the discussion of charges below. 

[Pro-Hyp-Gly]$_4$[Leu-Hyp-Gly][Pro-Hyp-Gly]$_5$: We also consider a collagen-like peptide relevant for biomineralization \cite{okuyama2007}, PDB entry 2drx, see Figure 7 c). Structures with hydroxyproline form a scaffold for mineralization, and the positions of the OH group of hydroxyproline become particularly important. In Figure 7 d) we show the CP structure, generated by the method described above. One observes that the OH groups of hydroxyproline sit in distinctively different patterns in the two structures. On the CP structure, the oxygens of hydroxyproline sit in two beautiful helical lines on the surface of the tropocollagen with about a 5 \AA ~repetition. The oxygen of the peptide units are internal in the tropocollagen and close to the channel. For the SC structure, the OH group of hydroxyproline sits in several non-equivalent sites on the surface intermixed with oxygens of the peptide units. For the CP structure, the pitch angle of the helical line formed by the hydroxyproline OH groups depends on the detailed values of the parameters of the structure and the depicted structure is as such a first estimate.  The oxygens of the peptides units point towards the channel effectively charging the inner surface of the channel. Presumably this negative charge is balanced by the presence of protons, see the discussion on NMR and the magic angle below. The inner surface is less charged if each of the three polypeptide strands are rotated around their own axes. Rotations are restricted as they must maintain the hydroxyproline OH groups accessible on the outer surface of the tropocollagen. 

\newsubsection{Biomineralization} 

Biomineralization is an important process which takes place in various kinds of life \cite{knoll2003}.  For tropocollagen, perhaps the most significant difference between the suggested close-packed and the supercoiled structure is the existence of a central channel in the CP structure not present in the SC-triple-helix. The diameter of the channel was estimated above to be $\sim$ 2 \AA , too small for water to be transported. However, it would be quite suitable for protons (ionic radius $<$0.01 \AA), for ionic phosphorus (ionic radius 0.3 \AA)  and perhaps for Ca$^{++}$ (ionic radius 1 \AA). We suggest that this is an essential feature enabling biomineralization by allowing for a separation of the positive and negative ions needed for the mineralization. In biomineralization phosphorus and calcium rich minerals are formed, e.g. hydroxyapatite, calcium carbonate and octacalcium phosphate. In bone formation, the hydroxyapatite mineral is formed in the space between the tropocollagen. 

It seems reasonable to assume that the nucleation of minerals takes place on the OH group of hydroxyproline and involves the about two layers thick hydration layer of water. Without a central channel to provide feeding of calcium and phosphorus ions these ions would have to diffuse through the hydration layer. The hydration layer is so thin (about two layers of H$_2$O) that it is barely thicker than the screening length of the ionic charges and the ionic groups. It is therefore quite possible that the various charges would grid-lock each other and thereby prevent ionic transportation and thus mineralization in taking place. Even if grid-lock would not persist, it seems peculiar that the collagen would mineralize in a uniform manner. As with nucleation in general, places where nucleation takes place first are likely to grow the largest grains of mineralized material. This would presumably destroy the ordered structure of the mesh formed by the collagen triple helices as well as of the fibrils. Instead, we suggest that the phosphorus and calcium ions are transported down the collagen triple helix through the central channel to the place where the mineralization takes place. Here mineralization continues radially outwardly until the ions cannot diffuse to the nucleation front. This may explains why mineralization takes place starting at one end of the triple collagen helix and working its way along the molecule; it also explains why the mineralization is only a few atomic layers thick filling the space between the collagen helices.

\newsubsection{NMR and the magic angle effect} 

In NMR studies of ordered collagen fibers, e.g. tendon, a striking effect (sometimes named the magic angle effect) has been observed \cite{berendsen1962,fullerton2007}. The effect is important for MR imaging of tendon \cite{marshall2002}.  The magic angle effect is the observation in NMR studies of two split proton-proton dipole interaction peaks which vanish at an angle of about $\epsilon=55^\circ$. 
At this angle, the second-order Legendre polynomial $P_2(\cos(\epsilon))=3 \cos^2 \epsilon -1$ vanishes. The results show that the dominant contribution stems from pairs of protons which are aligned parallel to the long axis of the tropocollagen. The data is consistent with a Gaussian distribution aligned with the axis, and with a width of the angular distribution of about 13$^\circ$ (the distribution of alignment of the tropocollagen molecules were not characterized and could contribute to the 13$^\circ$ spread) \cite{berendsen1962}. It seems plausible to us that the magic-angle effect can be due to protons being hosted by a central channel of the packed triple helix effectively forming a one-dimensional proton lattice. Previously, the signal has been suggested to originate from ordered chains of water molecules aligned to the collagen molecule such that the proton-proton vector is parallel with the direction of the tropocollagen \cite{berendsen1962,fullerton2007}. Curiously, this NMR phenomenon has not been observed for other fibrous molecules; in studies of silk fibroin, DNA, and keratin it was found that none of these fibrous molecules exhibit the magic angle effect  \cite{berendsen1965}.

\newsubsection{Circular dichroism}

Collagen has been extensively studied by measurements of circular dichroism (CD) spectra \cite{wallace2001}; nevertheless, the experimental data has not been fully described by model calculations. Bhatnagar and Gough writes "It is clear ... that the present theoretical methods do not adequately explain the CD of polyproline II and collagen" \cite{bhatnagar1996}. In short, both quantum mechanical exciton based calculation, and classical dipole based calculation fails to adequately describe the experimental data. For a related discussion with comparative results for different theoretical approaches, see Carlson et al. \cite{carlson2006}. Perhaps, the most straightforward explanation for this is that the current structural understanding of the collagen molecule is inadequate.

\newsection{Conclusion}

We have derived the condition for triple helices to be packed and calculated the volume fraction for the close-packed helix. The volume fraction comes about by defining an enclosing cylinder, and by considering its measure of volume as an indicator of how closely packed the three helices are. It is found that the condition for being close-packed is to have a pitch angle of $v_{CP} =43.3 ^\circ$. Table 1 summarizes these geometrical findings which holds for any triple helix of three tubes. Further, we have shown that there exists a zero-twist (ZT) structure with zero strain-twist coupling, which is a desirable molecular property for constituents of tendon and bone. For triple helices, the pitch angle of the two structures (CP and ZT) are found to be practically identical. Based on these findings we have suggested a new structure for collagen, which differs from the 10/3 and 7/2 conformations. Its three most fundamental features are those of being close-packed, of having zero strain-twist coupling,  and of having a central channel. It seems peculiar if Nature would not have taken advantage of this structure.

We find support for the existence of a central channel in early x-ray diffraction data of collagen.  The possibility for ion transport in the central channel suggests the need for a reexamination of various biological and physical phenomena, including biomineralization. 
The new conformation of collagen could also be relevant for understanding some of the collagen related diseases such as osteoporosis, Bethlem myopathy, arterial aneurysm and many more \cite{myllyharju2001}.

Some of the difficulties for the current understanding of the structure of collagen have been reviewed: the ambiguity in assigning crystal structures to crystallographic data for collagen-like peptides, and the failure to satisfactorily calculate circular dichroism data from model structures. Perhaps the angular dependence of the piezoelectric effect in bone (Fukada and Yasuda \cite{fukada1957}) with its maximum at $\sim 45^\circ$ is related to the pitch angle of the geometry of the hard wall tube model. For a description of the piezoelectric effect in helices, see Kasai  \cite{kasai1969}. Further, it is interesting to notice that Tiaho et al. have reported the determination of a pitch angle of about $41^\circ$ in a second harmonic generation imaging microscopy study \cite{tiaho2007}; a value which is close to the CP value. 

One of the distinctions between the different structures suggested for the tropocollagen molecule are the positions of the OH groups of hydroxyproline on the surface of the molecular structure. Experimental studies with atomic resolution of the mineralization of collagen are likely to be performed within the next 5-10 years, thereby helping to cast light on which of the features of the various suggested collagen structures (CP, 7/2, 10/3) are correct.

\end{document}